\documentclass[aps,twocolumn,pra]{revtex4-1}
\usepackage{graphics}
\usepackage{dcolumn}
\usepackage{bm}
\usepackage{amsmath}
\usepackage{hyperref}
\hypersetup{colorlinks=true, urlcolor=blue}
\usepackage{color}
\usepackage{tikz}
\usepackage{multirow}

\usepackage{float}
\usepackage[caption = false]{subfig}
\usepackage{newlfont}
\usepackage{amssymb}
\usepackage{amsfonts}
\usepackage{amsthm}
\usepackage{graphicx}
\usepackage{bm}

\def \be{\begin{equation}}
\def \ee{ \end{equation} }

\begin{document}

\definecolor{red}{rgb}{1,0,0}
\title{Simple Algorithms for Multi-term Bell-like Bases and Their Quantum Correlations}
\author{Asutosh Kumar}
\email{asukumar@hri.res.in}

\affiliation{Harish-Chandra Research Institute, Chhatnag Road, Jhunsi, Allahabad 211 019, India}

\begin{abstract}
We introduce two multiple qubit controlled-unitary gates with different working principles. We employ these gates and existing 
quantum gates to propose simple and efficient algorithms that generate multi-term orthonormal entangled Bell-like bases. All  algorithms thus far known for constructing entangled bases turn out to be special cases of our method. The Bell-like states in any basis is superposition of $2^m$-terms with equal probabilities (that is, their amplitudes being $\pm 1/\sqrt{2^m}$). The orthogonality of the basis does not permit arbitrary amplitudes. The quantum correlations of these bases are investigated; we find that the Bell-like bases obtained using different controlled-unitaries have different entanglement contents. We also learn that monogamy score is able to distinguish these bases in the situations where other quantum correlations fail to do so indicating that monogamy score is a more fundamental quantum correlation measure. Our approach can be extended to qudit systems as well.
\end{abstract}

\maketitle

\section{Introduction}
Entanglement is the characteristic trait of quantum mechanics \cite{schrodinger}, and quantum correlations \cite{horodecki09, modidiscord} are essential ingredients for performing quantum information tasks in both quantum communication and quantum computation. Highly genuinely entangled multiqubit states are the key resources of various quantum error correction codes and quantum communication protocols \cite{teleportation,densecoding,werner-pra65,cleve-prl83,gour-wallach}. 
Seeing vast applications of quantum correlations, entanglement in particular, in quantum information theory considerable attention is given to identification, generation, characterization, and quantification of quantum correlations. There have been attempts to find ways of generating optimal entanglement for given nonlocal interactions \cite{zanardi-pra62,dur-00,cirac-prl86}. In Ref. \cite{kraus-pra63} conditions for creating optimal entanglement using a two-qubit gate have been explored. Much efforts are being made to find maximally entangled states \cite{higuchi-sudbery00,brown05,brierley-higuchi07,borras07}, and construct entangled orthonormal bases. Entangled bases, e.g., two qubit Bell basis, are very important in quantum information tasks. For \(n\) qubit computational basis, 
$\{|x_1x_2\cdots x_n\rangle\}$, the two-term Bell basis and $2^n$-term Graph basis can be generated, respectively, using the expressions
\begin{align}
\label{eq:bellbasis}
|{\cal B}\rangle_{x_1x_2\cdots x_n} &=\frac{1}{\sqrt{2}}(|0x_2\cdots x_n\rangle +(-1)^{x_1}|1\bar{x}_2\cdots \bar{x}_n\rangle) \nonumber \\
&=C(X^{\otimes n-1})|(Hx_1)x_2\cdots x_n\rangle,
\end{align} 
and
\begin{equation}
\label{eq:graphbasis}
|{\cal G}\rangle_{x_1x_2\cdots x_n} =C(Z^{\otimes n-1})|H^{\otimes n}x_1x_2\cdots x_n\rangle, 
\end{equation}
where 
\begin{eqnarray}
  H=\frac{1}{\sqrt{2}}\left(
  \begin{array}{cc}
     1 & 1\\
     1 & -1
  \end{array}
  \right),
 X=\left(
  \begin{array}{cc}
     0 & 1\\
     1 & 0
  \end{array}
  \right),
 Z=\left(
  \begin{array}{cc}
     1 & 0\\
     0 & -1
  \end{array}
  \right),   
\end{eqnarray}
 and \(C(U)\) is a controlled-unitary operation. Both the bases are locally unitarily equivalent. For example,
 \begin{equation}
 |{\cal B}\rangle_{00}\xrightarrow{Z\otimes ZH}|{\cal G}\rangle_{11}.  
 \end{equation}
Very recently, an algorithm has been prescribed in Ref. \cite{braidbasis} to obtain $2^{n-1}$-term orthonormal entangled basis for \(n\)-qubit quantum systems exploiting the Braid theories, and the quantum properties of these states were analyzed. It was shown that any \(n\)-qubit Bell state in the basis has maximal concurrence and any two-qubit reduced density matrix is separable with zero concurrence. For instance, three-qubit and four-qubit bases using the Braid theories can be obtained as follows
\begin{equation}
 |x_1x_2x_3\rangle \longrightarrow (R\otimes I)(I\otimes R)|x_1x_2x_3\rangle,  
\end{equation}
and
\begin{align}
 |x_1x_2x_3x_4\rangle \longrightarrow & (R\otimes I\otimes I)(I\otimes R\otimes I) \nonumber \\ 
 &(I\otimes I\otimes R)|x_1x_2x_3x_4\rangle,  
\end{align}
where \(R\) is a unitary operator \cite{yangbaxteroperator}
\begin{eqnarray}
  R=\frac{1}{\sqrt{2}}\left(
  \begin{array}{cccc}
     1 & 0 & 0 & 1\\
     0 & 1 & -1 & 0\\
     0 & 1 & 1 & 0\\
     -1 & 0 & 0 & 1
  \end{array}
  \right)=
 \frac{1}{\sqrt{2}}\left(
  \begin{array}{cc}
     I & i\sigma_y\\
     i\sigma_y & I
  \end{array}
  \right)
\end{eqnarray}
Though this approach is straight forward for arbitrary \(n\)-qubit quantum systems, the computation becomes quite tedious for large \(n\). A natural question then arises: \textit{are there algorithms which are computationally simple and generate $2^m$-term, 
($1\leq m \leq n$), orthonormal entangled bases for \(n\)-qubit quantum systems?} 

In this paper we address this question. For that we first introduce two multiple qubit controlled-unitary operations with different working principles and then using these gates with existing multiqubit controlled-unitary operation and single qubit quantum gates we provide simple algorithms that generate $2^m$-term, ($1\leq m \leq n$), Bell-like orthonormal entangled bases for arbitrary \(n\)-qubit quantum systems. In any basis, \(n\)-qubit Bell-like states contain $2^m$ computational terms with equal probabilities (their amplitudes being $\pm 1/\sqrt{2^m}$); the orthogonality of the basis can not allow arbitrary amplitudes. We find that our approach is very general and all the known methods of obtaining entangled bases are contained in our approach. Besides this our approach can be extended to higher dimensional quantum systems as well.  
Several useful and readily computable quantum correlations 
like concurrence \cite{concurrence}, entanglement of formation (EoF) \cite{eof,concurrence}, logarithmic-negativity (LN) \cite{vidal-werner}, generalized geometric measure (GGM) \cite{ggm} (cf. \cite{gm}), quantum discord \cite{hv,oz}, quantum work-deficit \cite{work-deficit}, and monogamy score \cite{monogamyscore} (cf. \cite{monogamy}) of bipartite quantum correlation measures.
EoF is given in terms of concurrence \cite{concurrence} as given below
\begin{equation}
\mathcal{E}(\rho_{AB})=h\left(\frac{1+\sqrt{1-\mathcal{C}^2(\rho_{AB})}}{2}\right),
\label{eq:eof}
\end{equation}
where $h(x)=-x\log_2x-(1-x)\log_2(1-x)$ is the Shannon (binary) entropy. Negativity is half the value of concurrence for pure states.
GGM quantifies \emph{genuine} entanglement of multipartite pure systems. 
A multiparty pure quantum state is said to be genuinely multiparty entangled if it is entangled across every bipartition of its constituent parties.
A state is said to be maximally entangled if its average bipartite entanglement with respect to all possible bi-partitions is maximal (see \cite{love07,scott04} and references therein). 
Hence to quantify the global or average entanglement content in an \(n\)-party pure state, we compute the average entanglement of all possible bi-partite cuts given by
\begin{equation}
\langle{\cal Q}\rangle=\frac{1}{2^{n-1}-1}\sum_r {\cal Q}(\rho_r),
\end{equation}
where ${\cal Q}$ is some bipartite entanglement measure (viz., linear entropy, von Neumann entropy, negativity etc.) and $r \leq [\frac{n}{2}]$. 
Monogamy score can be interpreted as residual quantum correlation of the bi-partition $1:rest$ of an \(n\)-party state that cannot be accounted for by the quantum correlations of two-qubit reduced density matrices separately.
The investigation of these quantum correlations for Bell-like bases reveal interesting properties. 
The Bell-like bases obtained using different controlled-unitaries have, in general, different entanglement values.
For the Bell-like bases generated using one of the controlled-unitary operations we prove analytically that concurrence (hence EoF) and LN is unity.  We also learn that monogamy score can distinguish the Bell-like bases in the situations where other quantum correlations fail to do so indicating that monogamy score is a fine-grained quantum correlation measure. 

This paper is divided into four sections.  
In Sec. II, we introduce two multiqubit controlled-unitary gates. In Sec. III, we propose simple algorithms for constructing multi-term Bell-like bases, provide few explicit examples, and investigate quantum correlation properties of these bases. 
Finally, we conclude in Sec. IV.


\section{New Quantum Gates}
\label{gates}
Controlled-unitary operations can be defined with different working principles. In this section we enrich quantum operations by introducing two multiple qubit controlled-unitary operations with their working principles different from that of conventional controlled-U operation. Both are non-local quantum operations because their actions are defined on multiple qubits. Before describing them we briefly review some single-qubit and two-qubit quantum gates. 

The identity matrix, $\sigma_0\equiv I$, and the Pauli matrices $\{\sigma_x\equiv X, \sigma_y\equiv Y, \sigma_z\equiv Z\}$ when operate on single-qubit computational basis $\{|0\rangle,|1\rangle\}$ do, in order, nothing, bit-flip, bit-phase-flip, and phase-flip. Hadamard gate, $H=(X+Z)/\sqrt{2}$, on the other hand creates superposition of single qubit computational states. Another important two-qubit quantum gate is controlled-NOT (CNOT) gate whose action on two-qubit computational state $|x,y\rangle$ is described by
\begin{equation}
|x,y\rangle \xrightarrow{CNOT} |x,x\oplus y\rangle,
\end{equation}
where $x\oplus y$ denotes \emph{modulo 2 addition}, \(|x\rangle\) acts as control qubit and remains unchanged while \(|y\rangle\) acts as target qubit and is flipped when \(x=1\) only. In matrix form CNOT gate is represented as
\begin{eqnarray}
  CNOT=\left(
  \begin{array}{cc}
     I & 0\\
     0 & X
  \end{array}
  \right)
\end{eqnarray}
It has been shown that single qubit and CNOT gates are \emph{universal} \cite{divincenzo,barenco95}, i.e., any unitary operation can be approximated to arbitrary accuracy by a combination of CNOT and single qubit operations. Nonetheless, gates acting on multiple qubits can simplify the implementation of complex quantum algorithms \cite{chiaverini}. The multiqubit quantum operations can replace an intricate sequence of single and two-qubit gates, which in turn promises its apace execution with potentially higher fidelity. The quantum Toffoli gate, an important three-qubit control-control-not (CCNOT) gate, has been experimentally realized in NMR \cite{toffoli-nmr} and with ion traps \cite{toffoli-iontrap}. Moreover, circuit implementation of a quantum gate and its working principle are two different things. 

Consider a situation where conditioning is on multiple qubits and there are several target qubits. Suppose there are \(n+k\)
qubits, the first \(n\)-qubits are conditioned (i.e., act as control qubits), and \(U\) is a \(k\) qubit unitary operator. Then the controlled operation $C^n(U)$ is defined by the equation \cite{nielsen}
\begin{equation}
C^n(U)|x_1x_2\cdots x_n\rangle|\psi\rangle= |x_1x_2\cdots x_n\rangle U^{x_1x_2\cdots x_n}|\psi\rangle
\end{equation}
where $x_1x_2\cdots x_n$ in the exponent of \(U\) is the ordinary product of bits $x_1$, $x_2$, $\cdots$, $x_n$.
That is, the operator \(U\) is applied to the last \(k\) qubits if the first \(n\)-qubits are all equal to one, and do nothing otherwise. For \(n=k=1\) and \(U=X\), this is CNOT gate. For \(k\geq 2\) we don't yet know how to perform arbitrary operations on 
\(k\) qubits. The $C^n(U)$ operation is very important in quantum computations. 
Let us call this operation ``A1 or All1 (all one)" controlled-U operation because \(U\) is applied when all control qubits are equal to one. For instance, when $n=2$, $k=1$, and $U=X$ we have 
\begin{equation}
\label{eq:toffoli}
C^2(X)=diag \{I, I, I, X\}.
\end{equation}
Eq. (\ref{eq:toffoli}) represents the quantum Toffoli gate.

We now define the action of multiple qubit phase gate $\tilde{Z}$, which can be viewed as multiqubit controlled-\(Z\) operation, on an \(n\)-qubit computational state $|x_1x_2\cdots x_n\rangle$ as given below
\begin{align}
\tilde{Z}|x_1x_2\cdots x_n\rangle &= C^{n-1}(Z)|x_1x_2\cdots x_n\rangle \nonumber \\
&= (-1)^{x_1x_2\cdots x_n}|x_1x_2\cdots x_n\rangle
\end{align}
where $x_1x_2\cdots x_n$ in the exponent of -1 is the ordinary product of bits $x_1$, $x_2$, $\cdots$, $x_n$. That is, when all the qubits are one $\tilde{Z}$ gate will introduce a factor of -1, and do nothing otherwise. In matrix form $\tilde{Z}$ can be represented as
\begin{equation}
\label{eq:multiphase}
\tilde{Z}=diag \{I,I,\cdots,I,Z\}.
\end{equation}
From Eq. (\ref{eq:multiphase}) it is evident that $\tilde{Z}$ is unitary.\\

\subsection{\emph{odd one} controlled-unitary}
\label{oddcu}
Suppose there are \(n+k\) qubits where \(n\)-qubits $\{|x_1\rangle,|x_2\rangle,\cdots, |x_n\rangle\}$ are control qubits, \(k\) qubits $\{|y_1\rangle,|y_2\rangle,\cdots, |y_k\rangle\}$ are target qubits, and \(U_{r_j}\) is an \(r_j\)-qubit unitary operator with $1\leq r_j\leq k$ such that $\sum_j r_j=k$. That is, \(U_{r_1}\) acts on first \(r_1\) target qubits, \(U_{r_2}\) acts on next \(r_2\) target qubits, and so on. Let $U=\{U_{r_j}\}$. Then we define the action of the controlled operation $\tilde{C}_{O1}^n(U)$ as given below
\begin{align}
\tilde{C}_{O1}^n(U)|x_1x_2\cdots x_n\rangle |y_1y_2\cdots y_k\rangle=& |x_1x_2\cdots x_n\rangle \otimes \nonumber \\
& U^{x_c}|y_1y_2\cdots y_k\rangle,
\end{align}
where $x_c$ in the exponent of \(U\) is given by
\begin{equation}
\label{eq:xcontrol}
x_c=x_1\oplus x_2\oplus \cdots \oplus x_n
\end{equation} 
denotes \emph{modulo 2 addition}. That is, the operators $\{U_{r_j}\}$ are applied to $\{r_j\}$ qubits as described above if \(x_c=1\) (that is, when the number of 
$|1\rangle$'s is odd among control qubits), and do nothing if \(x_c=0\) (that is, when the number of 
$|1\rangle$'s is even among control qubits). For obvious reasons we dub this operation as ``O1 or Odd1 (odd one)" controlled-U gate. 
(To further complicate this gate one can devise \(U_{r_j}\)'s to operate in any desired order and/or operate on some common target qubits.) For \(n=k=1\) and \(U=X\), this is again the usual CNOT gate. But for \(n\geq 2\), $\tilde{C}_{O1}^n(U)$ is characteristically different from $C^n(U)$.  
For instance, when $n=2$, $k=1$, and $U=X$ we have 
\begin{equation}
\tilde{C}_{O1}^2(X)=diag \{I, X, X, I\}.
\end{equation}

\subsection{\emph{all equal} controlled-unitary}
\label{allequal-controlled-unitary}
Suppose there are \(n+k\) qubits, where the first \(n\)-qubits are control qubits, and \(U\) is a \(k\)-qubit unitary operator. We define the action of the controlled-U operation $\tilde{C}_{AQ}^n(U)$ by the equation 
\begin{equation}
\tilde{C}_{AQ}^n(U)|x_1x_2\cdots x_n\rangle|\psi\rangle=
|x_1x_2\cdots x_n\rangle U^{x'_c}|\psi\rangle
\end{equation}
where $x'_c$ in the exponent of \(U\) is equal to one when all the control qubits are equal, and zero otherwise.
That is, the operator \(U\) is applied to the last \(k\) qubits if the first \(n\) control qubits are all equal, and do nothing otherwise.
We refer to this controlled-U operation as ``AQ or AllQ (all equal)" controlled-U gate. This gate is valid only when the number of control qubits exceeds \(2\), and   is characteristically different from $C^n(U)$ and $\tilde{C}_{O1}^n(U)$.  
For $n=2$, $k=1$, and $U=X$ we have 
\begin{equation}
\tilde{C}_{AQ}^2(X)=diag \{X, I, I, X\}.
\end{equation}

\section{Algorithms for Bell-like Bases and Their Quantum Correlations}
\label{bellbasis}
In this section we describe several algorithms that are capable of generating $2^m$-term, ($1\leq m \leq n$), Bell-like orthonormal entangled bases for arbitrary \(n\)-qubit quantum systems, provide few explicit examples, and investigate their quantum correlation properties.  

\subsection{Algorithms}
For constructing $2^m$-term, ($1\leq m \leq n$), Bell-like orthogonal entangled bases for arbitrary \(n\)-qubit quantum systems, following three steps have to be performed on the computational basis $\{|x_1x_2\cdots x_n\rangle\}$:\\

(I) Apply Hadamard operation (\(H\)) on any \(m\) qubits and promote these qubits as control qubits. This creates a superposition of $2^m$ computational terms. For convenience we will consider the first \(m\) qubits as control qubits.\\

(II) Apply any one of the contolled-U operations on remaining (\(n-m\)) qubits to each term in (I): (CA1) $C^m(U)$, (CO1) $\tilde{C}_{O1}^m(U)$, and 
(CAQ) $\tilde{C}_{AQ}^m(U)$. \textit{When Hadamard operation is performed on \(m\) qubits the controlled-U operations $C^m(U)$, $\tilde{C}_{AQ}^m(U)$, and $\tilde{C}_{O1}^m(U)$ allow \(U\) to be applied once, twice, and $2^{m-1}$ times respectively}.
The phase operations, discussed below in (III), may or may not be associated with these controlled-U operations. Throughout this paper, we will take \(U\) to be single-qubit flip gate (\(X\)). Then
\begin{align}
\tilde{C}_{O1}^n(X^{\otimes k})|x_1\cdots x_n\rangle |y_1\cdots y_k\rangle=&
 |x_1\cdots x_n\rangle \otimes \nonumber \\
 &|\otimes^k_{j=1}(x_c\oplus y_j)\rangle,
\end{align}
where \(x_c\) is given by Eq. (\ref{eq:xcontrol}), and
\begin{align}
\tilde{C}_{AQ}^n(X^{\otimes k})|x_1\cdots x_n\rangle |y_1\cdots y_k\rangle=&
|x_1\cdots x_n\rangle \otimes \nonumber \\
 &|\otimes^k_{j=1}(X^{x'_c} y_j)\rangle,
\end{align}
where $x'_c$ is equal to one when all the control qubits are equal, and zero otherwise.\\

(III) Apply any one of the phase operations to each term in (I):
(P0) $(-1)^0=1$ (do nothing),
(P1) $(-1)^{x_1+\cdots +x_m}$ (apply \(Z\) gate on each control qubit),
(P2) $(-1)^{x_1x_2+x_2x_3+\cdots +x_mx_1}$,
(P3) $(-1)^{x_1x_2x_3+x_3x_4x_5+\cdots +x_{m-1}x_mx_1}$,$\cdots$, 
(Pm) $(-1)^{x_1x_2\cdots x_m}$. 

The phase operations (P2) through (Pm) can be realized using $\tilde{Z}$ gate. It is emphasized here that the phase operations (P0) through (Pm) do not rule out other possibilities. One can engineer numerous such phase transformations. For instance, one can exploit all the qubits of computational states for phase operations, the control qubits for phase operation and controlled-U operation can be different, there can be hierarchy among control qubits, and so on. An example is the phase operation (Pz) $C(Z^{\otimes m-1})$, in which the first qubit \(x_1\) acts as control qubit while \(Z\) gate is applied on qubits \(x_2\) through \(x_m\). \textit{While the phase operations (P0) and (P1) are locally equivalent, the phase operations (P2) and (Pz) yield the same phase factors for \(m=2\)}.\\

\begin{table}[ht]
\begin{tabular}{|c|c|}
\hline
 $\longrightarrow$ & Bell-like state                        \\ \hline \hline
$|000\rangle$ & \multicolumn{1}{|c|}{$|000\rangle + |010\rangle + |100\rangle + |111\rangle$} \\ \hline
$|001\rangle$ & $|001\rangle + |011\rangle + |101\rangle + |110\rangle$                       \\ \hline
$|010\rangle$ & $|000\rangle - |010\rangle + |100\rangle - |111\rangle$                       \\ \hline
$|011\rangle$ & $|001\rangle - |011\rangle + |101\rangle - |110\rangle$                       \\ \hline
$|100\rangle$ & $|000\rangle + |010\rangle - |100\rangle - |111\rangle$                       \\ \hline
$|101\rangle$ & $|001\rangle + |011\rangle - |101\rangle - |110\rangle$                       \\ \hline
$|110\rangle$ & $|000\rangle - |010\rangle - |100\rangle + |111\rangle$                       \\ \hline
$|111\rangle$ & $|001\rangle - |011\rangle - |101\rangle + |110\rangle$                       \\ \hline
\end{tabular}
\caption{Three-qubit four-term orthogonal unnormalized Bell-like basis $(3,2,CA1,P0)$ generated using controlled-U operation (CA1) and phase operation (P0). Here 
$|000\rangle \xrightarrow{(3,2,CA1,P0)}|000\rangle + |010\rangle + |100\rangle + |111\rangle$.}
\label{table:32P0CA1}
\end{table} 
 
\begin{table}[ht]
\begin{tabular}{|c|c|}
\hline
 $\longrightarrow$ &  Bell-like state                      \\ \hline \hline
$|000\rangle$ & \multicolumn{1}{|c|}{$|000\rangle + |011\rangle + |101\rangle + |110\rangle$} \\ \hline
$|001\rangle$ & $|001\rangle + |010\rangle + |100\rangle + |111\rangle$                       \\ \hline
$|010\rangle$ & $|000\rangle - |011\rangle + |101\rangle - |110\rangle$                       \\ \hline
$|011\rangle$ & $|001\rangle - |010\rangle + |100\rangle - |111\rangle$                       \\ \hline
$|100\rangle$ & $|000\rangle + |011\rangle - |101\rangle - |110\rangle$                       \\ \hline
$|101\rangle$ & $|001\rangle + |010\rangle - |100\rangle - |111\rangle$                       \\ \hline
$|110\rangle$ & $|000\rangle - |011\rangle - |101\rangle + |110\rangle$                       \\ \hline
$|111\rangle$ & $|001\rangle - |010\rangle - |100\rangle + |111\rangle$                       \\ \hline
\end{tabular}
\caption{Three-qubit four-term orthogonal unnormalized Bell-like basis $(3,2,CO1,P0)$ generated using controlled-U operation (CO1) and phase operation (P0). Here 
$|000\rangle \xrightarrow{(3,2,CO1,P0)}|000\rangle + |011\rangle + |101\rangle + |110\rangle$.}
\label{table:32P0CO1}
\end{table}

Since the phase operations (P0) through (Pm) and (Pz) are determined using only control qubits, the steps (II) and (III) are independent of each other and can be performed in any order.
For a particular basis, the phase operation and the controlled-U operation should be fixed.
Thus a particular basis can be denoted as $(n,m,Cq,Pp)$ where $q\in \{A1,O1,AQ\}$, and $p\in \{0,1,\cdots,m,z\}$. One might have noticed by now that once the Hadamard operation \cite{asu-hadamard} is done on the control qubits of the computational basis, it is very easy to generate the entangled Bell-like bases. It should be noted that the Bell basis in Eq. (\ref{eq:bellbasis}) and the graph basis in Eq. (\ref{eq:graphbasis}) can be obtained using the above approach.
It can be easily checked that all the Bell-like states in any basis are locally unitarily equivalent. On the contrary, two Bell-like bases may not be locally unitarily equivalent. 

We wish to emphasize here that the ways of obtaining orthonormal entangled Bell-like bases as addressed above can also be extended to higher dimensional quantum systems. One can define a large number of phase operations and controlled-unitary operations, sometimes complex, with different working principles.   
 
To illustrate, we tabulate the basis $(3,2,CA1,P0)$ in Table \ref{table:32P0CA1}, the basis $(3,2,CO1,P0)$ in Table \ref{table:32P0CO1}, and the bases $(3,2,CO1,P2)$, $(3,2,CAQ,P2)$ and the three-qubit basis obtained using operator \(R\) in Table \ref{table:32P2/PzCO1/R}. 
From Table \ref{table:32P2/PzCO1/R} we see that the bases $(3,2,CO1,P2)$ and $(3,2,CAQ,P2)$ are identical. Also the Bell-like states in the basis $(3,2,CO1,P2)$, and the states obtained using unitary operator \(R\) are equivalent upto a global factor -1 and permutation of computational states. \emph{This observation is also true for arbitrary \(n\) when \(m=n-1\) and the phase operation is (P2) (see Table \ref{table:43CO1P2/Braid} for another illustration)}. Thus our algorithm has obvious advantage over the Braid theory method for obtaining orthonormal entangled basis.

\subsection{Quantum Correlations of Bell-like Bases}
\label{qcorrelations}
We now investigate quantum correlations of the Bell-like bases. Firstly, we prove analytically that the concurrence of Bell-like states obtained using controlled-unitary (CO1) is unity.\\

\textbf{Proposition:} \textit{The concurrence of Bell-like states in the basis $(n,m,CO1,Pp)$ is unity.}\\

\texttt{Proof.} A Bell-like state in the basis $(n,m,CO1,Pp)$ is 
\begin{align}
|\psi^{CO1}_{x_1\cdots x_n}\rangle =& \tilde{C}_{O1}^m(X^{\otimes n-m})(-1)^*
|\otimes^m_{i=1}(Hx_i)\rangle|x_{m+1}\cdots x_n\rangle \nonumber \\
=&\sum_{y_1,\cdots ,y_m}(-1)^*|y_1\cdots y_m\rangle |\otimes^n_{j=m+1}(y_c\oplus x_j)\rangle \nonumber \\
=&\sum_{y_1,\cdots ,y_m}(-1)^*|y_1\rangle |y_2\cdots y_m\otimes^n_{j=m+1}(y_c\oplus x_j)\rangle,
\end{align}
where $y_c=y_1\oplus y_2\oplus \cdots \oplus y_m$ and $(-1)^*$ denotes some phase operation. Since no two product states 
$|y_2\cdots y_m\otimes^n_{j=m+1}(y_c\oplus x_j)\rangle$ are identical in the Bell-like state $|\psi^{CO1}_{x_1\cdots x_n}\rangle$ (see Tables \ref{table:32P0CO1},  \ref{table:32P2/PzCO1/R} and \ref{table:43CO1P2/Braid} for instances), we obtain $\rho_1=\mbox{tr}_{\bar{1}}(|\psi\rangle^{CO1}_{x_1\cdots x_n}\langle \psi|)=I/2$. Hence ${\cal C}(|\psi^{CO1}_{x_1\cdots x_n}\rangle)=2\sqrt{det \rho_1}=1$.
\hfill $\blacksquare$\\

Consequently, $S_1\equiv S(\rho_1)=1$, entanglement of formation, 
${\cal E}(|\psi^{CO1}_{x_1\cdots x_n}\rangle)=1$ using 
Eq. (\ref{eq:eof}), and since negativity is half of the concurrence for pure states, logarithmic-negativity \cite{vidal-werner} 
\begin{equation}
E_{\cal N}(|\psi^{CO1}_{x_1\cdots x_n}\rangle) = \log_2 [2 {\cal N}
(|\psi^{CO1}_{x_1\cdots x_n}\rangle) + 1]=1.
\label{eq:LN}
\end{equation}

\begin{widetext}
\begin{center}
\begin{table}[ht]
\begin{tabular}{|c|c|c|c|}
\hline
 $(3,2,CO1,P2)$ & $(3,2,CAQ,P2)$ & $(3,2,Braid)$ & Bell-like state   \\ \hline \hline
$|000\rangle$ & \multicolumn{1}{|c|}{$|001\rangle$} & $|011\rangle$ & $|000\rangle + |011\rangle + |101\rangle - |110\rangle$  \\ \hline
$|001\rangle$ & $|000\rangle$ & $|001\rangle$ & $|001\rangle + |010\rangle + |100\rangle - |111\rangle$                        \\ \hline
$|010\rangle$ & $|011\rangle$ & $|101\rangle$ & $|000\rangle - |011\rangle + |101\rangle + |110\rangle$                        \\ \hline
$|011\rangle$ & $|010\rangle$ & $|111\rangle$ & $|001\rangle - |010\rangle + |100\rangle + |111\rangle$                        \\ \hline
$|100\rangle$ & $|101\rangle$ & $|110\rangle$ & $|000\rangle + |011\rangle - |101\rangle + |110\rangle$                        \\ \hline
$|101\rangle$ & $|100\rangle$ & $-|100\rangle$ & $|001\rangle + |010\rangle - |100\rangle + |111\rangle$                        \\ \hline
$|110\rangle$ & $|111\rangle$ & $|000\rangle$ & $|000\rangle - |011\rangle - |101\rangle - |110\rangle$                        \\ \hline
$|111\rangle$ & $|110\rangle$ & $-|010\rangle$ & $|001\rangle - |010\rangle - |100\rangle - |111\rangle$                        \\ \hline
\end{tabular}
\caption{Three-qubit four-term bases $(3,2,CO1,P2)$ and $(3,2,CAQ,P2)$ generated using controlled-U operations (CO1) and (CAQ) respectively, and phase operation (P2), and the basis obtained using unitary operator \(R\) in the Braid theories. Here
$|000\rangle \xrightarrow{(3,2,CO1,P2)} |000\rangle + |011\rangle + |101\rangle - |110\rangle$, $|001\rangle \xrightarrow{(3,2,CAQ,P2)} |000\rangle + |011\rangle + |101\rangle - |110\rangle$, and $|011\rangle \xrightarrow{(3,2,Braid)} |000\rangle + |011\rangle + |101\rangle - |110\rangle$. Other rows have similar interpretations. 
Its apparent that all three bases are equivalent, that is, orthogonal Bell-like states (here shown unnormalized) in all three bases are same upto a global factor -1 and permutation of computational states.}
\label{table:32P2/PzCO1/R}
\end{table} 

\begin{table}[ht]
\begin{tabular}{|c|c|c|}
\hline
$(4,3,CO1,P2)$ & $(4,3,Braid)$                    & Bell-like state                                                                                                        \\ \hline \hline
$|0000\rangle$ & \multicolumn{1}{|c|}{$|0011\rangle$} & $|0000\rangle + |0011\rangle + |0101\rangle - |0110\rangle + |1001\rangle - |1010\rangle - |1100\rangle -|1111\rangle$ \\ \hline
$|0001\rangle$ & $|0001\rangle$                       & $|0001\rangle + |0010\rangle + |0100\rangle - |0111\rangle + |1000\rangle - |1011\rangle - |1101\rangle -|1110\rangle$ \\ \hline
$|0010\rangle$ & $|0101\rangle$                       & $|0000\rangle - |0011\rangle + |0101\rangle + |0110\rangle + |1001\rangle + |1010\rangle - |1100\rangle +|1111\rangle$ \\ \hline
$|0011\rangle$ & $|0111\rangle$                       & $|0001\rangle - |0010\rangle + |0100\rangle + |0111\rangle + |1000\rangle + |1011\rangle - |1101\rangle +|1110\rangle$ \\ \hline
$|0100\rangle$ & $|1111\rangle$                       & $|0000\rangle + |0011\rangle - |0101\rangle + |0110\rangle + |1001\rangle - |1010\rangle + |1100\rangle +|1111\rangle$ \\ \hline
$|0101\rangle$ & $|1101\rangle$                       & $|0001\rangle + |0010\rangle - |0100\rangle + |0111\rangle + |1000\rangle - |1011\rangle + |1101\rangle +|1110\rangle$ \\ \hline
$|0110\rangle$ & $|1001\rangle$                       & $|0000\rangle - |0011\rangle - |0101\rangle - |0110\rangle + |1001\rangle + |1010\rangle + |1100\rangle -|1111\rangle$ \\ \hline
$|0111\rangle$ & $|1011\rangle$                       & $|0001\rangle - |0010\rangle - |0100\rangle - |0111\rangle + |1000\rangle + |1011\rangle + |1101\rangle -|1110\rangle$ \\ \hline
$|1000\rangle$ & $|1010\rangle$                       & $|0000\rangle + |0011\rangle + |0101\rangle - |0110\rangle - |1001\rangle + |1010\rangle + |1100\rangle +|1111\rangle$ \\ \hline
$|1001\rangle$ & $-|1000\rangle$                       & $|0001\rangle + |0010\rangle + |0100\rangle - |0111\rangle - |1000\rangle + |1011\rangle + |1101\rangle +|1110\rangle$ \\ \hline
$|1010\rangle$ & $|1100\rangle$                       & $|0000\rangle - |0011\rangle + |0101\rangle + |0110\rangle - |1001\rangle - |1010\rangle + |1100\rangle -|1111\rangle$ \\ \hline
$|1011\rangle$ & $-|1110\rangle$                       & $|0001\rangle - |0010\rangle + |0100\rangle + |0111\rangle - |1000\rangle - |1011\rangle + |1101\rangle -|1110\rangle$ \\ \hline
$|1100\rangle$ & $|0110\rangle$                       & $|0000\rangle + |0011\rangle - |0101\rangle + |0110\rangle - |1001\rangle + |1010\rangle - |1100\rangle -|1111\rangle$ \\ \hline
$|1101\rangle$ & $-|0100\rangle$                       & $|0001\rangle + |0010\rangle - |0100\rangle + |0111\rangle - |1000\rangle + |1011\rangle - |1101\rangle -|1110\rangle$ \\ \hline
$|1110\rangle$ & $|0000\rangle$                       & $|0000\rangle - |0011\rangle - |0101\rangle - |0110\rangle - |1001\rangle - |1010\rangle - |1100\rangle +|1111\rangle$ \\ \hline
$|1111\rangle$ & $-|0010\rangle$                       & $|0001\rangle - |0010\rangle - |0100\rangle - |0111\rangle - |1000\rangle - |1011\rangle - |1101\rangle +|1110\rangle$ \\ \hline
\end{tabular}
\caption{Four-qubit eight-term basis $(4,3,CO1,P2)$ generated using controlled-U operations (CO1) and phase operation (P2), and the basis obtained using unitary operator \(R\) in the Braid theories \cite{braidbasis}. Here
$|0000\rangle \xrightarrow{(4,3,CO1,P2)} |0000\rangle + |0011\rangle + |0101\rangle - |0110\rangle + |1001\rangle - |1010\rangle - |1100\rangle -|1111\rangle$, and $|0011\rangle \xrightarrow{(4,3,Braid)} |0000\rangle + |0011\rangle + |0101\rangle - |0110\rangle + |1001\rangle - |1010\rangle - |1100\rangle -|1111\rangle$. Other rows have similar interpretations. 
Both bases are equivalent, that is, orthogonal Bell-like states (here shown unnormalized) in both bases are same upto a global factor -1 and permutation of computational states.}
\label{table:43CO1P2/Braid}
\end{table}
\end{center}
\end{widetext}

We also investigated numerically quantum correlations 
of orthonormal Bell-like bases $(n,m,Cq,Pp)$ for $n=3,4,5$; $1\leq m <n$; $q=O1,AQ,A1$; $p=0,1,\cdots ,m,z$. 
Since the Bell-like states in any basis are locally unitarily equivalent and any measure of entanglement is not changed by local unitary operations, all the Bell-like states in the basis have same entanglement content for the given measure. For pure bipartite quantum states discord and work-deficit are equal to von Neumann entropy of either of the reduced density matrix. Interestingly, information-theoretic measures like discord-score and work-deficit score also have fixed values for all the Bell-like states in a given basis obtained as described before (see Table \ref{table:measuresvalueCO1CAQCA1}). 
\textit{The phase operations, like controlled-unitaries, are also important. Two or more Bell-like bases with same \(n\), \(m\), and controlled-unitary can have different quantum correlation values for different phase operations (see Table \ref{table:measuresvalueCO1CAQCA1})}.

For all three controlled-U operations, values of quantum correlations are listed in Table \ref{table:measuresvalueCO1CAQCA1}. The bases obtained using these controlled-U operations often have identical values for different quantum correlation measures. For CO1-bases, in several instances where $\langle S \rangle$ is degenerate $\delta_{{\cal D}}$ is able to lift the degeneracy which indicates that monogamy score is a more fundamental quantum correlation measure. This is because two distinct sets of density matrices having same eigenvalue spectrum may not have the same set of discord values. Since ${\cal Q}=\delta_{{\cal Q}}=1$ (${\cal Q}={\cal C},{\cal E},E_{\cal N}$) and GGM has the maximal value \(0.5\), the Bell-like states obtained using controlled-unitary (CO1) are genuinely highly entangled and their two-qubit reduced density matrices are non-entangled. The high values of $\delta_{{\cal D}}$ indicate that two-qubit reduced density matrices are often classical-classical in nature. We found that discord score ($\delta_{{\cal D}}$) is equal to work-deficit score ($\delta_{\vartriangle}$) for CO1-bases.
For the controlled-U operation (CAQ) GGM varies as $\frac{1}{2^{m-1}}$, and for the controlled-U operation (CA1) GGM varies as $\frac{1}{2^m}$ irrespective of \(n\) and phase operations. Thus CO1-bases are the most genuinely entangled while CA1-bases are the least genuinely entangled for given value of \(m\). CO1-bases are also more entangled than CAQ- and CA1-bases. This establishes the effectiveness of CO1-unitary over other controlled-unitaries in entanglement generation.
 
\begin{widetext}
\begin{center}
\begin{table}[ht]
\begin{tabular}{|c|c|c|c|c|c|c|c|c|c|c|c|}
\hline
\multirow{2}{*}{$(n,m,Pp)$} & \multicolumn{3}{|c}{$\xi$} & \multicolumn{3}{|c}{${\cal C}$} & \multicolumn{3}{|c|}{$\langle S\rangle$} & $\delta_{{\cal C}}$ & $\delta_{{\cal D}}$ \\ \cline{2-12} 
                            & CO1   & CAQ     & CA1      & CO1   & CAQ        & CA1        & CO1         & CAQ          & CA1         & CO1                 & CO1                 \\ \hline \hline
$(n,1,P0)$                  & 0.5   & NA      & 0.5      & 1     & NA         & 1.0          & 1.0           & NA           & 1.0           & 1                   & 1.0                 \\ \hline
$(3,2,P0)$                  & 0.5   & 0.5     & 0.25     & 1     & 1.0          & 0.866025   & 1.0           & 1.0            & 0.811278    & 1                   & 0.994185            \\ \hline
$(3,2,P2)$                  & 0.5   & 0.5     & 0.25     & 1     & 1.0          & 0.866025   & 1.0           & 1.0            & 0.811278    & 1                   & 0.993259            \\ \hline
$(4,2,P0)$                  & 0.5   & 0.5     & 0.25     & 1     & 1.0          & 0.866025   & 1.4         & 1.4          & 0.976292    & 1                   & 0.997092            \\ \hline
$(4,2,P2)$                  & 0.5   & 0.5     & 0.25     & 1     & 1.0          & 0.866025   & 1.4         & 1.4          & 0.976292    & 1                   & 0.996629            \\ \hline
$(4,3,P0)$                  & 0.5   & 0.25    & 0.125    & 1     & 0.866025   & 0.661438   & 1.0           & 1.0588       & 0.61106     & 1                   & 0.991277            \\ \hline
$(4,3,P2)$                  & 0.5   & 0.25    & 0.125    & 1     & 0.866025   & 0.968246   & 1.0           & 1.0588       & 1.09811     & 1                   & 0.989888            \\ \hline
$(4,3,P3)$                  & 0.5   & 0.25    & 0.125    & 1     & 0.866025   & 0.661438   & 1.3         & 1.0588       & 0.61106     & 1                   & 0.0656589           \\ \hline
$(4,3,Pz)$                  & 0.5   & 0.25    & 0.125    & 1     & 0.866025   & 0.968246   & 1.4         & 1.24737      & 1.09811     & 1                   & 0.996629            \\ \hline
$(5,2,P0)$                  & 0.5   & 0.5     & 0.25     & 1     & 1.0          & 0.866025   & 1.4         & 1.4          & 0.976292    & 1                   & 0.997092            \\ \hline
$(5,2,P2)$                  & 0.5   & 0.5     & 0.25     & 1     & 1.0          & 0.866025   & 1.4         & 1.4          & 0.976292    & 1                   & 0.996629            \\ \hline
$(5,3,P0)$                  & 0.5   & 0.25    & 0.125    & 1     & 0.866025   & 0.661438   & 1.4         & 1.18879      & 0.688201    & 1                   & 0.994185            \\ \hline
$(5,3,P2)$                  & 0.5   & 0.25    & 0.125    & 1     & 0.866025   & 0.968246   & 1.4         & 1.18879      & 1.09328     & 1                   & 0.993259            \\ \hline
$(5,3,P3)$                  & 0.5   & 0.25    & 0.125    & 1     & 0.866025   & 0.661438   & 1.5         & 1.18879      & 0.688201    & 1                   & 0.377106            \\ \hline
$(5,3,Pz)$                  & 0.5   & 0.25    & 0.125    & 1     & 0.866025   & 0.968246   & 1.53333     & 1.33423      & 1.09328     & 1                   & 1.0                 \\ \hline
$(5,4,P0)$                  & 0.5   & 0.125   & 0.0625   & 1     & 0.661438   & 0.484123   & 1.0           & 0.747128     & 0.403766    & 1                   & 0.98837             \\ \hline
$(5,4,P2)$                  & 0.5   & 0.125   & 0.0625   & 1     & 0.968246   & 0.992157   & 1.53333     & 1.34022      & 1.29369     & 1                   & 0.997092            \\ \hline
$(5,4,P3)$                  & 0.5   & 0.125   & 0.0625   & 1     & 0.968246   & 0.927025   & 1.41634     & 1.27634      & 1.10593     & 1                   & 0.0654596           \\ \hline
$(5,4,P4)$                  & 0.5   & 0.125   & 0.0625   & 1     & 0.661438   & 0.484123   & 1.27043     & 0.747128     & 0.403766    & 1                   & 0.446504            \\ \hline
$(5,4,Pz)$                  & 0.5   & 0.125   & 0.0625   & 1     & 1.0          & 0.992157   & 1.4         & 1.24804      & 1.10757     & 1                   & 0.996629            \\ \hline
\end{tabular}
\caption{Values of GGM ($\xi$), concurrence (${\cal C}$), and average entanglement entropy ($\langle S \rangle$) of orthonormal Bell-like bases $(n,m,Cq,Pp)$ for $n=3,4,5$, $1\leq m <n$, $q=O1,AQ,A1$, $p=0,1,\cdots, m,z$. ``NA" stands for ``not applicable". For controlled-U operation (CO1) all the bases have maximal GGM, $\xi=0.5$, for controlled-U operation (CAQ) GGM varies as $\frac{1}{2^{m-1}}$, and for controlled-U operation (CA1) GGM varies as $\frac{1}{2^m}$ irrespective of \(n\) and phase operations. Thus CO1-bases are the most genuinely entangled while CA1-bases are the least genuinely entangled for given value of \(m\). CO1-bases are also more entangled than CAQ- and CA1-bases. The bases obtained using these controlled-U operations often have identical values of different quantum correlation measures. For CO1-bases concurrence score ($\delta_{{\cal C}}$) and discord score ($\delta_{{\cal D}}$) have also been evaluted. 
We see that in several instances where $\langle S \rangle$ is degenerate $\delta_{{\cal D}}$ is able to lift the degeneracy, indicating that information-theoretic monogamy score is a fine-grained quantum correlation measure.}
\label{table:measuresvalueCO1CAQCA1}
\end{table}
\end{center}
\end{widetext}

\section{conclusion \& discussion}
\label{sec:summary}
We introduced two multiqubit controlled-unitary gates and proposed a number of simple and efficient ways to construct multi-term orthonormal Bell-like bases. The approach is very general and incorporates all earlier known methods of constructing orthonormal entangled bases. The number of ways to obtain such bases increases exponentially with increasing system size. This reveals entanglement complexity of quantum systems and affirms the fact that entanglement is the characteristic trait of quantum mechanics. 

We proved analytically that concurrence (hence entanglement of formation and logarithmic-negativity) for the Bell-like bases obtained using controlled-U operation (CO1) and arbitrary phase operation is unity.
We found that the Bell-like bases obtained using ``Odd1" controlled-U operation are  genuinely highly entangled but fragile to particle loss. We also observed that  monogamy scores (information-theoretic ones) can be effectively used to distinguish these bases when other quantum correlations fail to do so. This indicates that monogamy score is a fine-grained quantum correlation measure. 
For given \(m\), CO1-bases are the most genuinely entangled while CA1-bases are the least genuinely entangled. CO1-bases are also more entangled than CAQ- and CA1-bases for given \(n\), \(m\), and the phase operation. 

Since the Bell-like basis $(n,n-1,CO1,P2)$, and the basis obtained using the Braid theories are equivalent upto a global factor -1 and permutation of computational states, this motivates to find increased applications of the Braid theories in quantum information theory \cite{genbraidbasis}. 
The algorithms discussed in this paper create entangled states from computational states. When we learn how to perform arbitrary unitary operations on multiple qubits, this approach will reveal further structure and complexity of quantum correlations, especially entanglement, of quantum systems systematically.
We believe that the multiqubit controlled-U gates which we have introduced in this paper will contribute significantly, like $C^n(U)$, in quantum computations, 
and the Bell-like bases obtained as addressed before will find potential applications in quantum information tasks. Our approach can be straight forwardly generalized for arbitrary dimensional quantum systems where a wide variety of similar, sometimes complex, controlled-unitary and phase operations can be defined and orthonormal entangled bases can be obtained.\\

\begin{acknowledgements}
AK acknowledges Debasis Mondal for useful discussions and suggestions. 
\end{acknowledgements}

\end{document}